\documentclass[twocolumn,noshowpacs,preprintnumbers,amsmath,amssymb,superscriptaddress]{revtex4-2}
\usepackage{graphicx}
\usepackage{dcolumn}
\usepackage{bm}

\usepackage{siunitx}
\usepackage{upgreek}

\begin{document}

\title{Cascade switching current detectors based on arrays of Josephson junctions}

\author{Roger Cattaneo}
\affiliation{Department of Physics, Stockholm University, AlbaNova University
Center, SE-10691 Stockholm, Sweden.}

\author{Artemii E. Efimov}
\affiliation{Department of Physics, Stockholm University, AlbaNova University
Center, SE-10691 Stockholm, Sweden.}
\affiliation{Department of Physics, University of Basel, 4056 Basel, Switzerland}

\author{Kirill I. Shiianov}
\affiliation{Department of Physics, Stockholm University, AlbaNova University
Center, SE-10691 Stockholm, Sweden.}

\author{Oliver Kieler}
\affiliation{Physikalisch-Technische Bundesanstalt, 38116 Braunschweig, Germany}

\author{Vladimir M. Krasnov}
\email{Vladimir.Krasnov@fysik.su.se}
\affiliation{Department of Physics, Stockholm University, AlbaNova University
Center, SE-10691 Stockholm, Sweden.}


\begin{abstract}

Cascade multiplication is a widely used approach to enhance the sensitivity of photon detectors. Although vacuum tube and semiconductor photomultipliers achieve high signal gain and enable single-photon detection in the optical range, their performance is constrained at lower frequencies by a significant work function (1–10 eV). Superconducting detectors overcome this limitation, enabling operation in the terahertz (THz) and microwave (MW) frequency ranges.
In this study, we introduce novel cascade-amplified superconducting detectors based on arrays of Josephson junctions. The interjunction coupling induces avalanche-like switching of multiple junctions upon photon absorption, resulting in cascade amplification of the readout voltage. We present two prototypes featuring either low-$T_c$ linear $\mathrm{Nb/Nb_{x}Si_{1-x}/Nb}$ arrays or high-$T_c$ stacked intrinsic Josephson junctions. Both MW and THz responses are analyzed and the advantage of the cascade detector compared to a conventional single-junction detector is demonstrated.  
Our findings suggest that array-based cascade superconducting detectors can be used to build broadband MW-to-THz sensors, with sensitivities exceeding $10^{13} \ \mathrm{V/W}$.
\end{abstract}

\maketitle

\section*{Introduction}

Detectors in the microwave (MW) and terahertz (THz) ranges find diverse applications, including security, environmental monitoring, medical imaging, chemical analysis, and fundamental research 
\cite{Zhang_2016,Lewis_2019,Eisman_2011}. Detectors with single-photon and photon-counting resolution are needed in quantum optics and electronics \cite{Eisman_2011,Inomata_2016,Kono_2018,Shevchenko_2024}. However, despite intense research in this area for the past two decades \cite{Zhang_2016,Lewis_2019,Eisman_2011,Zmuidzinas_2003,Gershenson_2008,Goltsman_2016,Inomata_2016,Kono_2018,AKuzmin_2018,Shevchenko_2024,Echarnach_2018,Govenius_2020,Shi_2022,Lau_2023,Krasnov_2024,Day_2024,Auton_2017,Kim_2021,Gaydchenko_2021,Astafiev_2000,Kajihara_2013,McDermott_2012,Ilichev_2017} MW and THz single-photon detectors (SPDs) are not yet commercially available due to many technical challenges and materials limitations. 

The main challenge is associated with small photon energies 
[0.41 - 41] meV for the frequency range [0.1-10] THz. 
This puts strong constraints on detector materials. Conventional vacuum tube and semiconductor photodetectors \cite{Eisman_2011} are not operational in this range due to the large work function and band gap (typically 1-10 eV). Therefore, low-gap materials such as superconductors \cite{Kono_2018,Zmuidzinas_2003,Gershenson_2008,Goltsman_2016,Inomata_2016,Echarnach_2018,AKuzmin_2018,Govenius_2020,Shevchenko_2024,Shi_2022,Lau_2023,Krasnov_2024,Day_2024}, half-metals \cite{Auton_2017,Govenius_2020,Kim_2021,Gaydchenko_2021}, or gap-engineered quantum dots \cite{Astafiev_2000,Kajihara_2013} are required. Furthermore, in situ gap tunability is needed for optimizing detector operation and adjusting dynamic range \cite{Krasnov_2024}. This can be achieved through electrostatic gating \cite{Kono_2018,Auton_2017,Echarnach_2018,Govenius_2020,Kim_2021,Gaydchenko_2021,Astafiev_2000,Kajihara_2013}, magnetic field \cite{Inomata_2016,Astafiev_2000,Krasnov_2024,Shevchenko_2024}, bias \cite{Goltsman_2016,Krasnov_2024,AKuzmin_2018,Shevchenko_2024,McDermott_2012,Ilichev_2017}, temperature \cite{Zmuidzinas_2003,Gershenson_2008,Goltsman_2016,AKuzmin_2018,Day_2024,Krasnov_2024,McDermott_2012,Ilichev_2017}, etc. A sensitive detector should have a high absorption efficiency, $\chi$, 
necessitating the implementation of a pickup antenna for impedance matching with free space 
\cite{Balanis,Krasnov_2023}.

The low photon energy also sets severe demands on detector characteristics. For MW SPDs, the noise-equivalent power (NEP) should be in the zW/Hz$^{1/2}$ range \cite{Astafiev_2000,Kajihara_2013,Echarnach_2018,Govenius_2020,Day_2024} (0.66 zW corresponds to absorption of one 1 THz photon per second).
So sensitive devices cannot operate at room temperature due to large black-body radiation with a power density of about $46$ mW/cm$^2$ and a peak frequency 17.6 THz at $T=300$ K. Therefore, cooling to cryogenic temperatures is commonly required, enabling the utilization of superconductors.  
Superconductivity is beneficial for building ultrasensitive detectors due to the lack of Jonson-Nyquist voltage noise, $S_V=[4k_bTR]^{1/2}$, in electrodes. Many of existing superconducting detectors, most noticeably those based on qubits \cite{Kono_2018,Inomata_2016,Echarnach_2018,Govenius_2020,Shevchenko_2024}, approach the quantum limit of sensitivity set by the Heisenberg uncertainty relation. The base element of qubits, exhibiting a quantum-mechanical behavior 
\cite{Martinis_1987}, is a Josephson junction (JJ). 

A current-biased JJ can act as a sensitive switching current detector (SCD) with potential for single-photon resolution in a broad MW to THz range \cite{Devyatov_1986,GrJensen_2004,McDermott_2012,Andersen_2013,Ilichev_2017,Borodianskyi_2017,Kuzmin_2020,Divin_2020,Cattaneo_2021,Torrioli_2022,Krasnov_2024}. 
The responsivity of SCD can be very high, limited only by quantum fluctuations
\cite{Krasnov_2024,Devyatov_1986}. 
The upper frequency 
depends on the characteristic voltage, $V_c = I_c R_n$, 
and ranges from sub-THz for low-$T_c$, 
to THz for high-$T_c$ \cite{Divin_2020,Katterwe_2012,Cattaneo_2021,Borodianskyi_2017} JJs. The highest $V_c \gtrsim 30\ \mathrm{mV}$  \cite{Krasnov_2002,Krasnov_2009} is achieved in intrinsic Josephson junctions (IJJ), which naturally occur in 
Bi$_2$Sr$_2$CaCu$_2$O$_{8+\delta}$ (Bi-2212) cuprates \cite{Kleiner_1994}. The operation of IJJs above 10 THz has been demonstrated \cite{Borodianskyi_2017,Katterwe_2012}. The atomic scale of IJJs enables strong mutual coupling, which is advantageous for the creation of coherent THz electronics
\cite{Ozyuzer_2007,Benseman_2013,Kashiwagi_2015,HBWang_2019,Ono_2020,Kakeya_2020,Borodianskyi_2017,Cattaneo_2021}. Mutual coupling between JJs leads to a current-locking phenomenon in JJ arrays \cite{Grebenchuk_2022}, where the switching of one JJ drags several neighbors into the resistive state, thereby cascade-multiplying the readout voltage \cite{Cybart_2019,Grebenchuk_2022,Golod_2019}.

\begin{figure*}[t]
    \centering
    \includegraphics[width=0.95\textwidth]{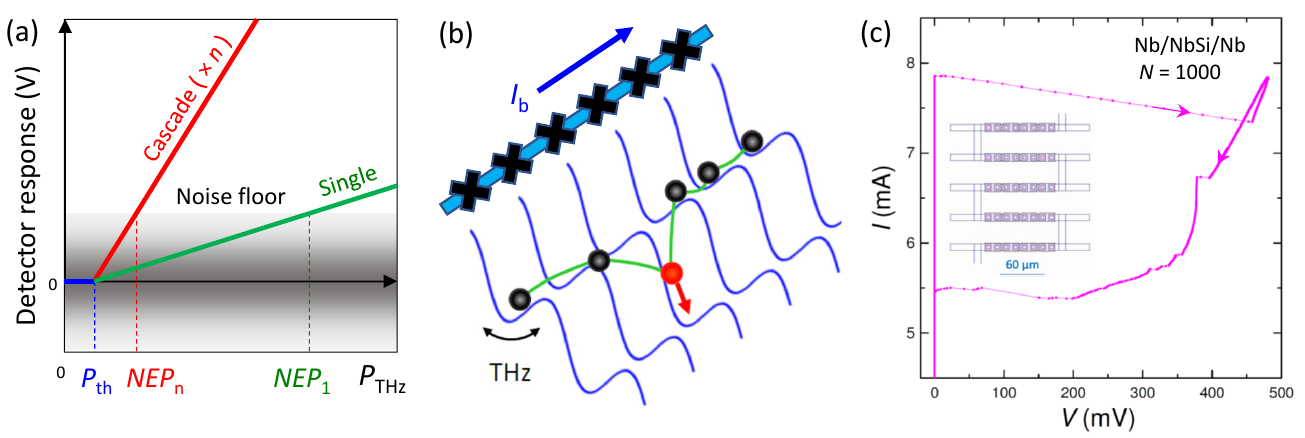}
    \caption{{\bf The concept of cascade amplification.} (a) Clarification of NEP reduction by cascade-amplification of the detector sensitivity. Olive and red lines indicate voltage responses of a single-stage and cascade-amplified detectors as a function of the incoming power. The gray shaded region indicates the readout voltage noise. (a) Energy-phase diagram of a current-biased JJ array. Switching (escape) of a single junction (red circle) leads to avalanche-like dragging of other JJs due to the mutual inter-junction coupling (green lines). (c) The $I$-$V$ characteristics of an array with $N=1000$ Nb/Nb$_x$Si$_{1-x}$/Nb JJs at $T\simeq 2.5$ K. It can be seen that all JJs switch simultaneously from the superconducting to the resistive state, leading to a large readout voltage, $V\simeq 0.45$ V. The inset shows the meander-like array layout.
    } 
    \label{fig:fig1}
\end{figure*}

In this work, we present a new concept of cascade-amplified SCDs based on arrays of coupled JJs. Collective, avalanche-like switching of many JJs leads to cascade amplification of the readout voltage, boosting the SCD sensitivity. We present two prototypes based on either linear arrays of low-$T_c$ Nb/Nb$_x$Si$_{1-x}$/Nb JJs, or stacked high-$T_c$ IJJs in Bi-2212 whiskers. We investigate both MW and THz responses and compare single-junction and cascade SCD operation at the same device. MW absorption efficiency is analyzed by studying polarization loss diagrams, revealing antenna effects associated with device geometry. We argue that the combination of high-$T_c$ superconductivity and large cascade gain can yield broadband THz sensors with outstanding sensitivities. 

\section*{Operation principle}

Cascade multiplication is widely employed to improve detector sensitivity, e.g., in photomultipliers and avalanche photodiodes \cite{Eisman_2011}. Figure \ref{fig:fig1} (a) clarifies the principle of a cascade-amplified detector. The green line shows the voltage response of a single-stage detector, as a function of the incoming THz power, $P_{THz}$. Above a threshold, $P_{THz}>P_{th}$, it grows linearly with the slope determined by the sensitivity, $S=V/P_{THz}$ (V/W). NEP corresponds to $P_{THz}$ at which the response exceeds the noise floor. The sensitivity of the cascade detector is increased in proportion to the number of stages, $S_n=n S_1$. This will reduce $NEP_n$, provided, the noise floor is not multiplied correspondingly.     
Our goal is to achieve cascade amplification of the readout voltage. Individual JJs inherently exhibit variations. Without interjunction coupling, the array $I$-$V$ is simply the sum of $I$-$V$'s of individual JJs and the response in the low-power limit is determined by the weakest (single) JJ. However, when the coupling is present, a current-locking phenomenon can occur \cite{Grebenchuk_2022}, causing all (or several) JJs to switch simultaneously, as sketched in Fig. \ref{fig:fig1} (b). Collective switching leads to a cascade multiplication of the readout voltage, proportional to the number of current-locked JJs \cite{Cybart_2019,Golod_2019,Grebenchuk_2022}. The efficiency of cascading depends on the uniformity of JJs and the strength of coupling. Therefore, arrays with many strongly interacting JJs are required. Fig. \ref{fig:fig1} (c) shows the $I$-$V$ of an array containing $N=1000$ Nb/Nb$_x$Si$_{1-x}$/Nb JJs \cite{Cattaneo_2022}.
It demonstrates a nearly ideal current locking with all JJs switch simultaneously, 
leading to a large readout voltage, $V\simeq 0.45$ V.

\begin{figure*}[t]
    \centering
    \includegraphics[width=0.95\textwidth]{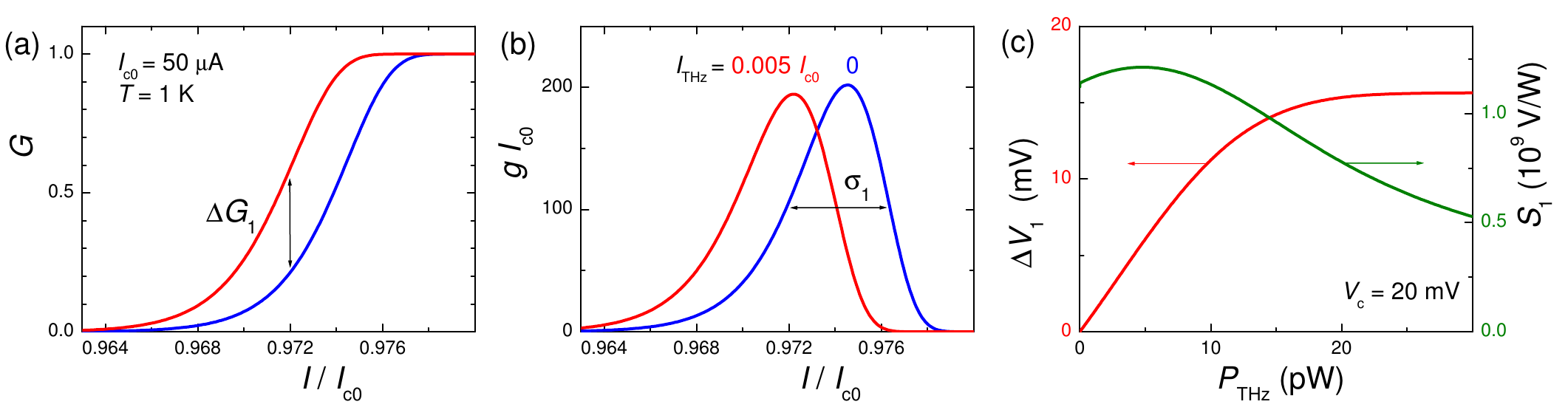}
    \caption{{\bf Operation of an ordinary single-junction SCD} (a) Calculated switching probabilities as a function of bias current for a JJ with $I_{c0}=50~\mu$A, $Q_0=100$ and $T=1$ K. The blue line - without radiation and the red line with induced THz current of $I_{THz}=0.005~I_{c0}$. (b) Corresponding probability densities, representing switching current histograms. (c) Calculated voltage response (red line, left axis) and sensitivity (olive line, right axis) versus incoming THz power for $I_b=0.972~I_{c0}$, as indicated by vertical arrows in (b). Based on the data from Ref. \cite{Krasnov_2024} for the non-resonant case, assuming $V_c=20$ mV, relevant for a single Bi-2212 IJJ.
    } 
    \label{fig:fig2}
\end{figure*}

\subsection*{Single-junction SCD}

Detailed description of a single-junction SCD can be found in Ref. \cite{Krasnov_2024}. The JJ dynamics is equivalent to motion of a particle in a tilted washboard potential, see Fig. \ref{fig:fig1} (b). The corrugation is determined by the Josephson energy $E_{J0}=(\Phi_0/2\pi)I_{c0}$, where $\Phi_0$ is the flux quantum and $I_{c0}$ is the fluctuation-free critical current; the tilt is created by the bias current, $I$; and the coordinate, $\varphi$, is given by the Josephson phase difference. When the particle escapes from the potential well, the JJ switches into the resistive state. This occurs at a switching current $I_s<I_{c0}$. The premature switching is caused both by the induced THz current from the impacting radiation, and by thermal and quantum current fluctuations. Therefore, SCD requires a statistical readout - the price which has to be paid for the exceptional sensitivity \cite{Krasnov_2024}. However, this is true for any quantum-limited detector.

The blue lines in Figure \ref{fig:fig2} (a) show a switching probability, $G(I)$, versus bias current in the absence of radiation. Calculations are made for an underdamped JJ with the quality factor $Q_0=100$, $I_{c0}=50~\mu$A at $T=1$ K \cite{Krasnov_2024}. 
The switching current histogram, $I_s(I)$, is given by the probability density, $g=dG/dI$, as shown in Fig. \ref{fig:fig2} (b). It has the full-width at half-maximum \cite{Jackel_1974},
\begin{equation}
\sigma_1 \simeq a I_{c0}\left[\frac{k_BT}{2E_{J0}}\right]^{2/3}.
    \label{Sigma1}
\end{equation}
According to our simulations, the prefactor, $a\sim 1$, slightly depends on $E_{J0}/k_B T$ and varies from $a=0.57$ for $E_{J0}/k_B T>100$ to $a\simeq 1$ for $E_{J0}/k_B T\sim 10$.

Red lines in Figs. \ref{fig:fig2} (a) and (b) show corresponding quantities in the presence of radiation-induced current with the amplitude, $I_{THz}=0.005 I_{c0}$. Calculations are done for the non-resonant case with the frequency much smaller than the Josephson plasma frequency \cite{Krasnov_2024}. The SCD is biased by a low-frequency alternating current with a constant amplitude $I_b$ close to the middle of the $G(I)$ step in the absence of radiation. The time-average voltage response $\Delta V$ is proportional to the vertical shift of $\Delta G(I_b)$ at the bias amplitude, as indicated by the arrows in Fig. \ref{fig:fig2} (a) 
\begin{equation}
    \Delta V_1 \simeq V_c \Delta G(I_b).
    \label{DeltaV1}
\end{equation}

The red line in Fig. \ref{fig:fig2} (c) shows SCD response, $V(P_{THz})$, at $I_b=0.972~I_{c0}$, marked in Fig. \ref{fig:fig2} (a), for $V_c=20$ mV, relevant for Bi-2212 IJJ. The response grows approximately linearly at small $P_{THz}$ and saturates when $I_{THz}$ exceeds $2\sigma_1$. The external $P_{THz}$ is connected to the internal $I_{THz}$ as, 
\begin{equation}
    P_{THz}=\frac{I_{THz}^2 R_{THz}}{2\chi}. 
    \label{P_THz}
\end{equation}
Here $\chi$ is the absorption efficiency and $R_{THz}$ is the real part of the THz impedance. Hereinafter, we consider the optimal impedance-matched case with maximum $\chi=0.5$ \cite{Balanis,Krasnov_2023}, assuming $R_{THz}=100~\Omega$ \cite{Siddiqi_2005}.

The olive line (right axis) in Fig. \ref{fig:fig2} (c) shows the corresponding single-JJ SCD sensitivity, $S_1$. It can be estimated using Eq. (19) from Ref. \cite{Krasnov_2024}. Taking into account that $\sigma_1\simeq 1/g_{max}$, where $g_{max}$ is the peak value of the histogram, 
we can write,
\begin{equation}
    S_1\sim 2\chi V_c R_{THz} \frac{e^2}{(\hbar \omega)^2} \frac{I_{c0}}{\sigma_1}.
    \label{S_1}
\end{equation}
For the chosen parameters, Eqs. (\ref{S_1}) and (\ref{Sigma1}) yield $S_1\simeq 10^9$ (V/W), consistent with Fig. \ref{fig:fig2} (c). However, the ultimate SCD sensitivity can be much higher, up to $S_1=5\times 10^{12}$ (V/W) \cite{Krasnov_2024}. From Eq. (\ref{S_1}) it is seen that the SCD sensitivity is limited only by the finite width of the switching current histogram. According to Eq.(\ref{Sigma1}), $\sigma_1$ reduces with decreasing $T$ due to the reduction of thermal fluctuations. However, it saturates below the crossover temperature to macroscopic quantum tunneling, $T^*\sim 10$ mK \cite{Martinis_1987}. Therefore, SCD represents probably the simplest quantum-limited detector. 

\subsection*{Multi-junction cascade SCD}

It has been known that JJ arrays can be advantageously used for photon detection: arrays can help with impedance matching and broaden the dynamic range \cite{Richards_1987,Koshelets_1991}. However, earlier studies were focusing on heterodyne mixers, operating at the quasiparticle branch of the $I$-$V$. In contrast, here we consider SCD, where the signal is generated upon switching out of the superconducting state 
\cite{McDermott_2012,Andersen_2013,Ilichev_2017,Kuzmin_2020,Krasnov_2024}.

The cascade amplification factor of the multi-junction SCD is given by the number of current-locked JJs, $n$, 
\begin{equation}
    V_n \simeq nV_1,~~~~(1\leq n \leq N).
    \label{Vn}
\end{equation}
Ideally, all $N$ junctions switch simultaneously, as shown in Fig. \ref{fig:fig1} (c). However, due to various types of nonuniformity this is not always the case. Therefore, the performance of array-based SCD depends on the distribution $n(P_{Thz})$ with the most important parameters being the starting number of JJs, $n(0)$, at $P_{THz}\rightarrow 0$ and the width of the distribution, $\sigma_n$.  

\begin{figure*}[t]
    \centering
    \includegraphics[width=0.95\textwidth]{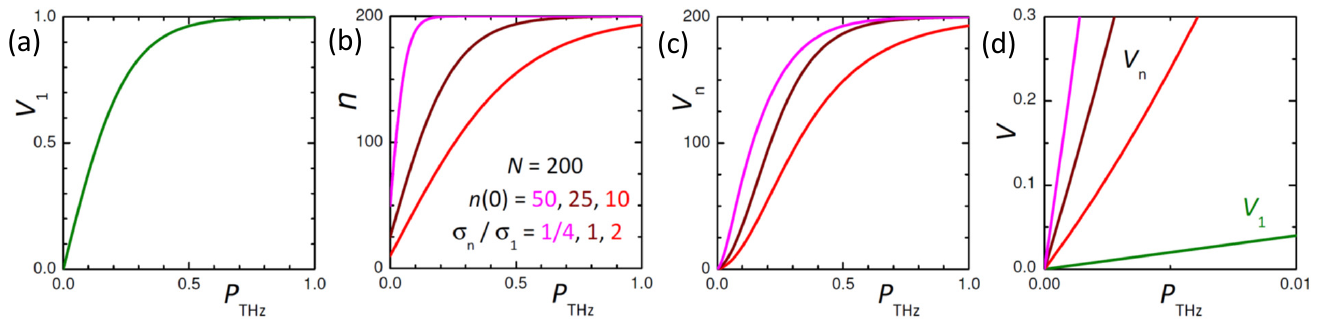}
    \caption{{\bf Calculation of sensitivity for a cascade SCD.} (a) A single junction response, characterized by the power-width $\sigma_1$. (b) Three examples of distribution functions of the number of active JJs versus $P_{THz}$ for an array with $N=200$ JJs. The leftmost (magenta) corresponds to a fairly uniform array with a large initial number $n(0)=50$ and a narrow power-width $\sigma_n=\sigma_1/4$. The middle (wine), to $n(0)=25$ and $\sigma_n=\sigma_1$. The rightmost (red), to the least uniform array with $n(0)=10$ and $\sigma_n=2\sigma_1$. Panels (c) and (d) show array responses, calculated from Eq. (\ref{Vn}), (c) in the full scale and (d) at small power.   } 
    \label{fig:fig3}
\end{figure*}

Figure \ref{fig:fig3} illustrates multi-junction SCD operation. Fig. \ref{fig:fig3} (a) shows the mean single-JJ response $V_1(P_{THz})$, characterized by the width $\sigma_1$, c.f. with the red line in Fig. \ref{fig:fig2} (c). Fig. \ref{fig:fig3} (b) represents three examples of possible $n(P_{THz})$ for arrays with $N=200$ JJs. The leftmost (magenta) shows the case of a fairly uniform array with $n(0)=50$ and with rapid activation of all JJs, $\sigma_n=1/4~\sigma_1$; the middle (wine) to $n(0)=25$ and $\sigma_n=\sigma_1$; and the rightmost (red) to the least homogeneous array with small $n(0)=10$ and broad $\sigma_n=2~\sigma_1$. Figs. \ref{fig:fig3} (c) and (d) represent corresponding array responses calculated via Eq. (\ref{Vn}), (c) in the full power range and (d) at low $P_{THz}$. It can be seen that larger $\sigma_n$ broadens the dynamic range of the detector, while the sensitivity at low power is determined solely by $n(0)$,
\begin{equation}
    S_n(0)=S_1 n(0).
\end{equation}

Additional clarifications about SCD operation, lock-in readout, statistical uncertainty and noise measurement are provided in Methods and Supplementary information.   

\section*{Experimental results}

Below we analyze MW and THz responses of low-$T_c$ Nb/Nb$_x$Si$_{1-x}$/Nb JJ arrays and high-$T_c$ mesa structures on Bi-2212 whiskers. 
Fabrication and physical properties of Nb arrays were described in Refs. \cite{Muller_2009,Galin_2018,Galin_2020,Cattaneo_2022}. Details about Bi-2212 device fabrication and characterization can be found in Refs. \cite{Krasnov_2013,Borodianskyi_2017,Cattaneo_2021} and the Supplementary. 
All presented measurements are performed at zero magnetic field and $T\simeq$ 3.3 K. Experimental details and additional data can be found in Methods and the Supplementary. 

\subsection*{Microwave detection}

A linearly polarized MW signal at $f\simeq 74.5 \ \mathrm{GHz}$  is delivered to the sample in a quasi-optical manner. 
MW attenuation, $\mu$, and polarization angle, $\Theta$, are adjusted using grid polarizers. The incoming MW power, $P$, is monitored by a Golay cell. 

\subsubsection*{Linear Nb arrays}

\begin{figure*}[t]
    \centering
    \includegraphics[width=0.95\textwidth]{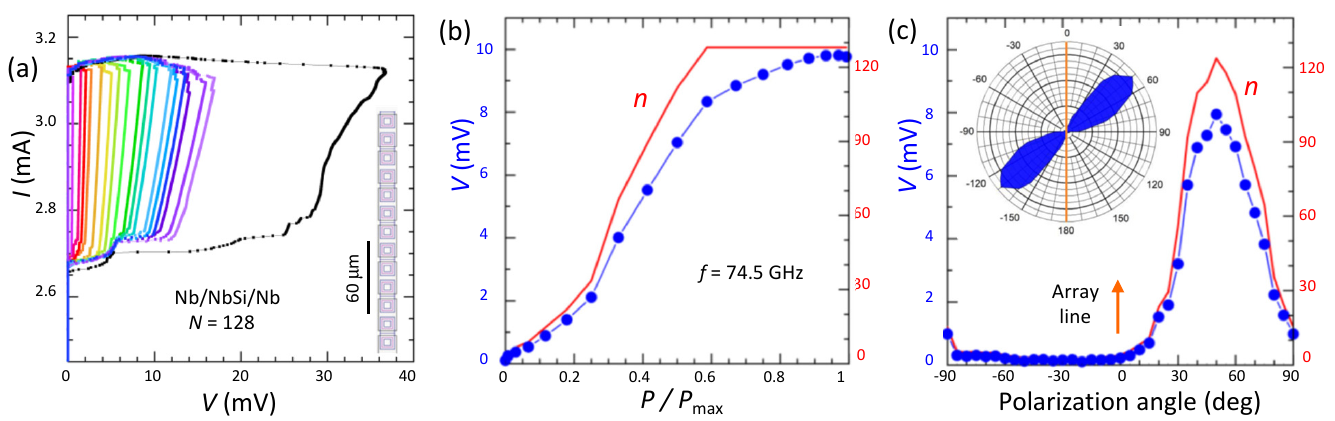}
 \caption{{\bf Cascade SCD based on a linear Nb/Nb$_x$Si$_{1-x}$/Nb array with $N=128$ JJs.} (a) Ensemble of the $I$-$V$'s without MW irradiation, obtained at slightly different bias currents.  
 The inset shows the array layout. 
    (b) The array response (blue symbols, left axis) and the number of active JJs, $n$, (red line, right axis) as a function of MW power. (c) $V$ and $n$ as a function of the polarization angle, $\Theta$, for $P=\text{const}$. The inset shows the polarization loss-function. 
The orange line marks the array direction. Profound off-axis lobes manifest the traveling-wave antenna effect.
    } 
    \label{fig:fig4}
\end{figure*}

We study a linear Nb array containing $N=128$ Nb/Nb$_{x}$Si$_{1-x}$/Nb JJs. 
Fig. \ref{fig:fig4} (a) shows the $I$-$V$s measured upon minor variation of the bias current in the absence of MW. 
Multiple branches correspond to a different number, $n$, of active JJs. 
Interjunction coupling in these arrays is mediated by surface plasmons \cite{Galin_2020} and is manifested by the appearance of collective resonant steps in the $I$-$V$'s \cite{Cattaneo_2022}. 

Blue circles in Fig. \ref{fig:fig4} (b) show the array response measured at a fixed $I_b$ 
as a function of MW power. With increasing $P$, the switching current, $I_s$, decreases and progressively more JJs switch into the resistive state, leading to the increase of readout voltage. The red line in Fig. \ref{fig:fig4} (b) represents the number of active JJs, $n$. It can be seen that the rapid increase of $V$ at low $P$ is proportional to $n$, illustrating the cascade amplification phenomenon. At higher power, all $N=128$ JJs switch to the resistive state and the responsivity $dV/dP$ is greatly reduced. In this case the response is solely due to the decrease of $I_s(P)$, qualitatively similar to a single-JJ SCD \cite{Borodianskyi_2017,Krasnov_2024}. 

As follows from Eq. (\ref{S_1}), detector sensitivity is directly proportional to the absorbtion efficiency, $\chi$. A JJ alone cannot effectively absorb radiation because its size is much smaller than the wavelength in free space, $\lambda_0$ \cite{Krasnov_2023}. However, the 2 mm long active part of the array is $\sim \lambda_0/2$ at 74.5 GHz. There are also passive electrodes (without JJ) with the total length $\sim 1\ \mathrm{cm}$. As shown in earlier studies of MW emission from such arrays \cite{Galin_2018,Galin_2020}, the long electrodes act as non-resonant traveling wave antennas, enabling impedance matching. The key signature of the traveling wave antenna is an asymmetric off-axis directivity \cite{Balanis,Galin_2018,Galin_2020}.

To analyze the absorption efficiency of the array, we measure its polarization loss factor \cite{IEEE}. 
Fig.\ref{fig:fig4} (c) shows the detector voltage (blue symbols, left axis) and the number of active JJs (red line, right axis) as a function of the MW polarization angle, $\Theta$, at a constant $P$. $\Theta=0$ corresponds to the MW electric field parallel to the array line. The inset displays the corresponding polarization-loss diagram, $\Delta V/V_{G}(\Theta)$, where $\Delta V$ is the array voltage with respect to the turned-off MW, and $V_{G}\propto P$ is the Golay cell voltage (constant in this experiment). 
The diagram exhibits two lobes with profound maxima at $\Theta \simeq 50^{\circ}$ and $-130^{\circ}$ with respect to the array. 
The off-axis behavior is consistent with traveling-wave antenna operation of long electrodes, enabling good impedance matching and absorption efficiency.   

\begin{figure*}[t]
    \centering
    \includegraphics[width=0.95\textwidth]{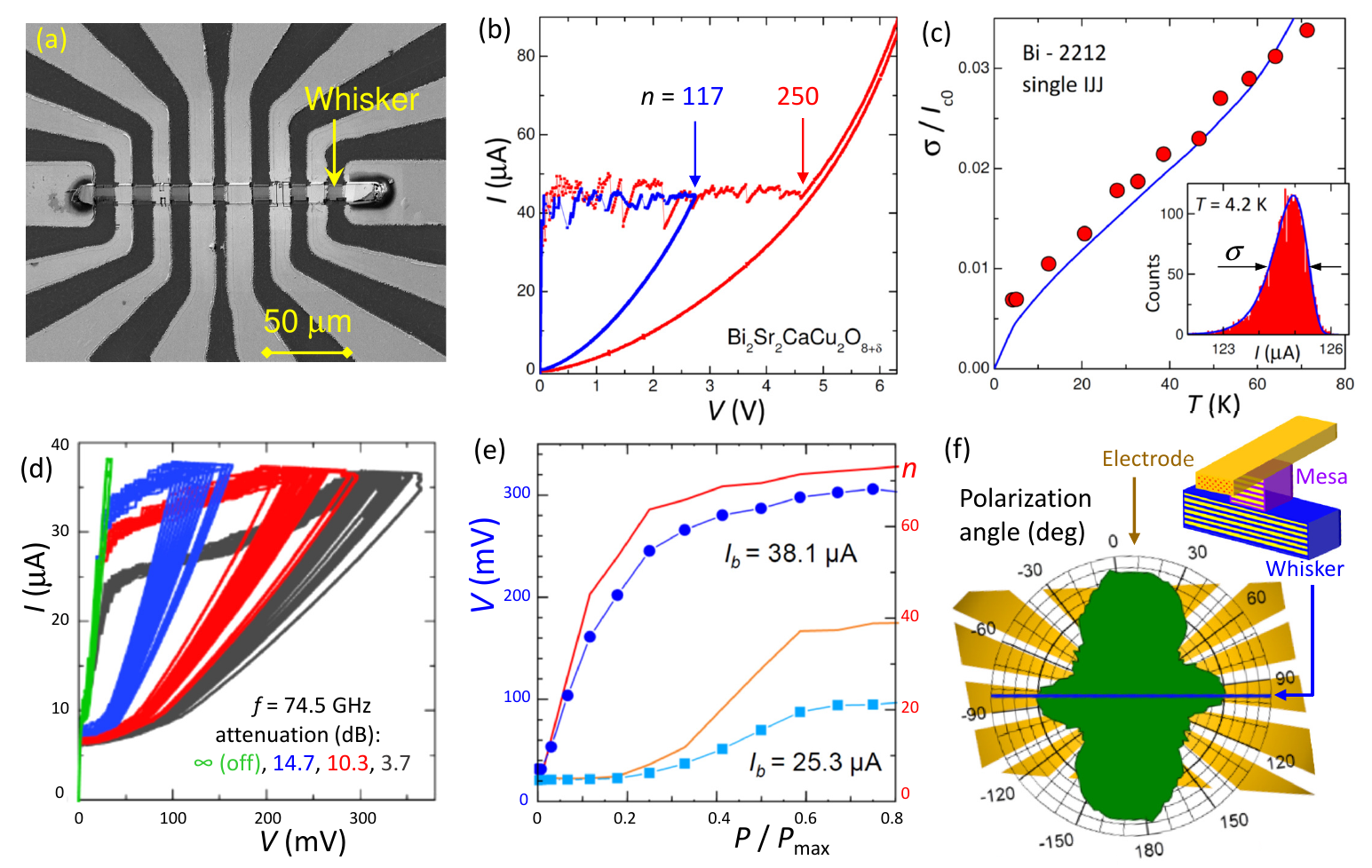}
\caption{ {\bf Microwave characteristics of a Bi-2212 mesa}. (a) A SEM image of a whisker-based device. 
(b) $I$-$V$s of a mesa with $N=250$ IJJs, measured with two different bias amplitudes without MW.
(c) Red circles represent measured $T$-dependence of the width of switching current histograms for a single IJJ. The blue line is calculated from Eq. (\ref{Sigma1}). The inset shows the measured switching current histogram (red) and the calculated switching probability density (blue line).  
(d) Ensemble of the $I$-$V$s for different MW powers, {\color{black} determined by the attenuation factor $\mu$}, at constant $I_b=38.1~\mu\mathrm{A}$. (e) The mesa voltage (symbols) and the number of active IJJs, $n$, (lines) versus the MW power, for $I_b=25.3$ and $38.1~\upmu\mathrm{A}$. (f) The {\color{black} polarization-loss}  diagram. The blue line indicates the orientation of the whisker. The yellow background shows the large-scale electrode geometry. The four-fold symmetry is consistent with the turnstile-antenna geometry of the device, as sketched in the inset.
}
       \label{fig:fig5}
\end{figure*}

\subsubsection*{Mesa structures on Bi-2212 whiskers}

Figure \ref{fig:fig5} (a) shows SEM image of one of the studied high-$T_c$ devices. They are based on Bi-2212 whiskers - needle like single crystals \cite{Cattaneo_2021}. 
Each device contains several mesas, representing stacks of IJJs. Mesas are formed at the intersection of the whisker with top gold electrodes, as sketched in Fig. \ref{fig:fig5} (f). To reduce the $I_c$ 
and increase the sensitivity, some mesas were trimmed 
using focused ion beam (FIB) etching \cite{Krasnov_2013,Borodianskyi_2017}. 

Fig. \ref{fig:fig5} (b) displays two $I$-$V$'s of a mesa ($\sim 5\times5$ $\upmu \mathrm{m}^2$), containing $N\simeq 250$ IJJs, without MW. The blue curve, with $n=117$ active IJJs, is measured with $I_b$ slightly above the mean switching current, $I_s \sim 50~\upmu\mathrm{A}$. 
A multi-branch structure ends at $V\simeq 5$ V when all $N\simeq 250$ IJJ's switch into the resistive state (red curve). The large readout voltage (5 V) is caused by the large $V_c \simeq 20 \ \mathrm{mV}$  per IJJ. Switching of the first and the last IJJ occurs within an interval, $\sigma_n \sim 5\ \upmu\mathrm{A}$ $\ $$\sim 0.1~I_s$, representing the width of the switching current distribution \cite{Krasnov_2005,Krasnov_2024}. 
Statistical analysis of multi-junction switching can be found in the Supplementary.

Inset in Fig. \ref{fig:fig5} (c) shows a switching current histogram (red) for a single IJJ on another mesa. The solid line shows the simulated probability density at the base $T$. Red circles in Fig. \ref{fig:fig5} (c) show measured temperature dependence of the histogram width. The blue line is obtained from Eq. (\ref{DeltaV1}). The overall agreement is excellent, indicating that IJJs are well described by the standard formalism.  

Fig. \ref{fig:fig5} (d) shows the $I$-$V$s of the third mesa 
at different MW powers and $I_b=38.1~\mu$A. The cascade gain 
depends on $I_b$. Fig. \ref{fig:fig5} (e) represents the MW power dependencies of $V$ (symbols) and $n$ (lines), measured at two biases. At lower $I_b = 25.3\ \upmu\mathrm{A}$, the sensitivity, $dV/dP$, remains small up to some threshold, $P \sim 0.2~P_{max}$. At higher $I_b = 38.1~\upmu\mathrm{A}$, there is no threshold and the sensitivity is high at $P=0$. With further increase of $I_b$, the sensitivity remains high at $P=0$ but saturates at successively smaller $P$, thus reducing the dynamic range \cite{Krasnov_2024}. Therefore, there is an optimal bias ($I_b = 38.1\ \upmu\mathrm{A}$ in this case) with both high sensitivity 
and large dynamic range. For comparison, in the Supplementary we analyze the ordinary SCD operation on the same mesa using a single surface IJJ. 
The surface IJJ has much smaller $I_s$ \cite{Krasnov_2009} and can be measured without activation of other IJJs \cite{Borodianskyi_2017}. The maximum single-junction response is almost 1000 times 
smaller than 
for the cascade SCD in Fig. \ref{fig:fig5} (e) at the same MW power for the same mesa.  

\begin{figure*}[t]
    \centering
    \includegraphics[width=0.95\textwidth]{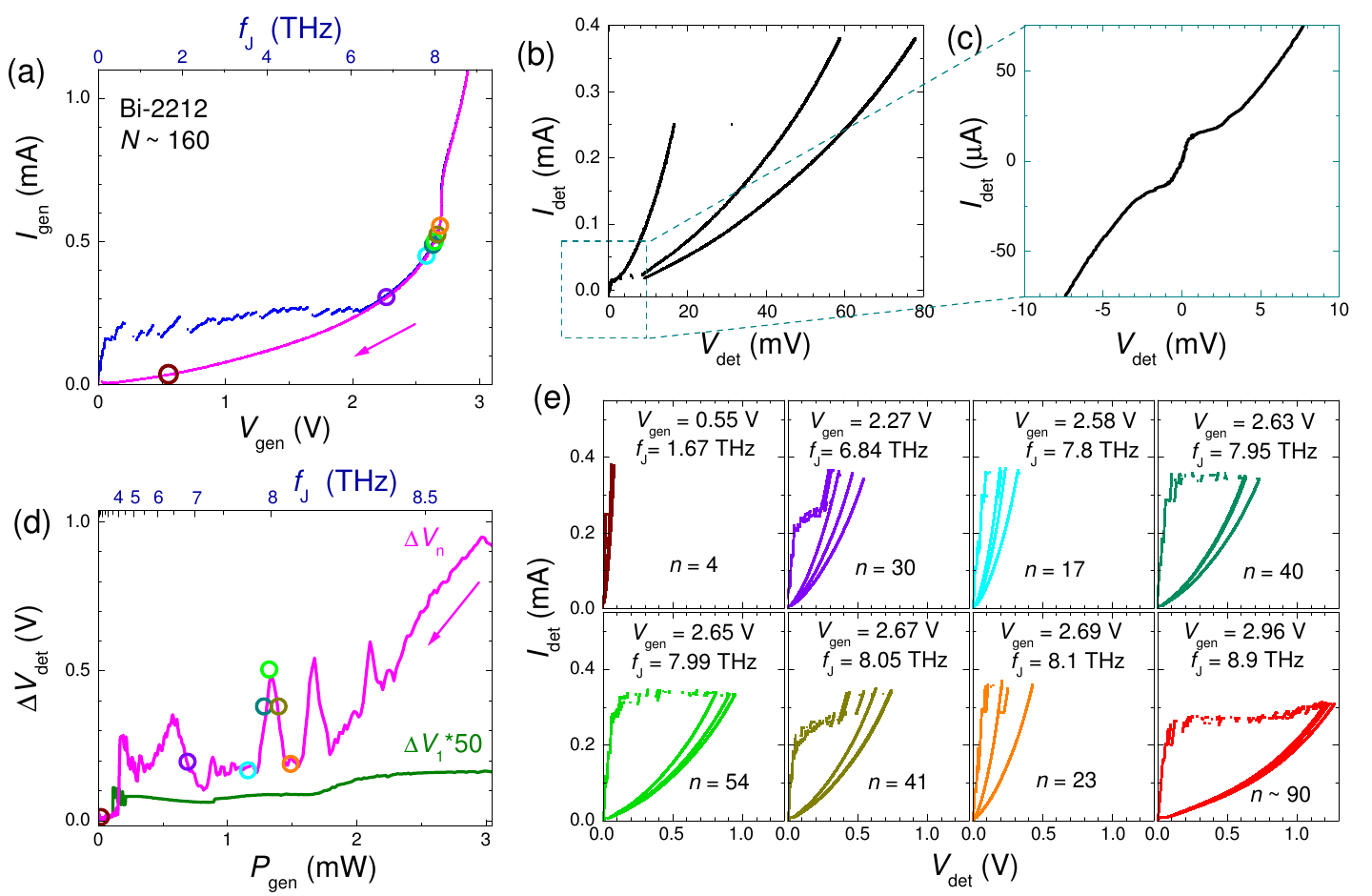}
\caption{ {\bf Detection of THz radiation from a Bi-2212 mesa}. (a) The $I$-$V$ 
of the generator mesa 
for upward (blue) and downward (magenta) bias sweeps. Top axis indicates the anticipated Josephson frequency. (b) and (c) Low-bias $I$-$V$'s of the detector mesa with (c) showing the $I$-$V$ of the weak surface junction. (d) Measured detector responses versus $P_{gen}=I_{gen}V_{gen}$. The olive line represents a single surface IJJ response (multiplied by a factor 50). The magenta curve represents the cascade SCD response of the same mesa. (e) A set of detector $I$-$V$'s with $I_{b}=370~\mu$A at different bias points of the generator, marked by the same-color circles in (a) and (d). 
}
       \label{fig:fig6}
\end{figure*}

Fig. \ref{fig:fig6} (f) shows the polarization-loss diagram of this mesa. 
It has a four-fold shape with two smaller lobes aligned with the whisker (indicated by the blue line) and the larger lobes approximately perpendicular, slightly inclined in the direction of one of bias electrodes (see the Supplementary for more details). The large-scale electrode geometry of this device is shown in the background (yellow). The observed four-fold diagram is consistent with the turnstile-antenna geometry of the device, as sketched in the inset \cite{Note1}.

\subsection*{Detection of THz radiation from a Bi-2212 mesa}

To verify THz operation, we perform an in-situ generation-detection experiment in which one mesa emits Josephson radiation and a nearby mesa detects it \cite{Borodianskyi_2017,Cattaneo_2021}. Fig. \ref{fig:fig6} (a) shows the $I$-$V$ of a generator mesa ($\sim 10\times 15~\mu$m$^2$), 
containing $N=160\pm 10$ IJJs. We analyze the downturn part of it (magenta) at which all $N$ IJJs are active. In this case, the Josephson frequency is well defined, $f_J=2eV/hN$, as indicated in the top axis. The emission occurs at cavity modes in the mesa \cite{Borodianskyi_2017,HBWang_2019,Ono_2020}. The key signature of such emission (as opposed to heating) is the non-monotonous dependence of the signal on the dissipation power, $P_{\text{gen}}= I_{\text{gen}} V_{\text{gen}}$, with distinct peaks at cavity resonances \cite{Borodianskyi_2017}. 

Fig. \ref{fig:fig6} (b) shows a low-bias part of the $I$-$V$ of a detector mesa ($\sim 15\times 15~\mu$m$^2$), about $22~\mu$m away from the generator mesa. Fig. \ref{fig:fig6} (c) is a close-up on the central part, representing the $I$-$V$ of the surface junction. It has $I_c' \simeq 15~\mu$A, significantly smaller than $I_c\simeq 370~\mu$A for the rest of the IJJs. This disparity allows ordinary SCD measurements at $I_c'<I_b\ll I_c$, which activates only one surface IJJ \cite{Borodianskyi_2017}. The cascade SCD can be analyzed on the same mesa by increasing bias to $I_{b}\sim I_c$. 

Fig. \ref{fig:fig6} (d) shows single-junction (olive, $\times 50$) and cascade (magenta) SCD responses versus $P_{gen}$, measured at bias amplitudes, 
$I_b=25$ and 370 $\mu$A, respectively. It can be seen that (in this case) the resolution of a single-junction SCD is not sufficient for resolving the emission. The cascade SCD demonstrates a much larger response with clearly visible emission peaks. The main emission for this generator mesa occurs in the range from $\sim 4$ to 8.4 THz, as indicated in the top axis. 
Fig. \ref{fig:fig6} (e) shows the detector $I$-$V$'s measured with $I_{b}=370~\mu$A at several bias points of the generator, marked by the same-color circles in Figs. \ref{fig:fig6} (a) and (d). It can be seen that cascade amplification factor, $n$, is playing the dominating role in the detector response. 

\section*{Discussion}

We observed 
similar behavior both for low- and high-$T_c$ devices, either linear arrays or stacks. Both types of detectors have some advantages. 2D-linear arrays facilitate a large absorption area and simple scalability to tens of thousands of JJs \cite{Muller_2009}. 

The list of advantages of high-$T_c$ devices include: 
(i) Broad frequency range covering the whole THz area [0.1-10] THz \cite{Borodianskyi_2017}.
(ii) Compactness in combination with large $N$.
(iii) Large $V_c \sim 20-30$ mV \cite{Krasnov_2002,Krasnov_2009}, enabling high readout signals. From Figs. \ref{fig:fig4} (a) and \ref{fig:fig5} (b) it can be seen that the maximum readout voltage, $NV_c$, for a high-$T_c$ mesa ($~5\ $V) is more than 100 times larger than for the low-$T_c$ array ($\sim 0.04$ V) despite only a factor two difference in $N$ (250 vs. 128). 
(iv) The atomic separation of IJJs enables the strongest interjunction coupling, needed for the avalanche switching. 

\subsection*{Sensitivity of a cascade SCD}

As described in the Supplementary, for Bi-2212 detector,  {\color{black} the signal, $\Delta V$,} increases from {\color{black} 0.32 mV} for the single junction SCD (no amplification, $n=1$), {\color{black} to 81 mV} at sub-optimal bias, and {\color{black} almost up to 0.3 V} at the optimal bias, Fig. \ref{fig:fig5} (e). We emphasize that these measurements were performed under the same conditions with the only difference in $I_b$, determining the cascade amplification.


The sensitivity $S$(V/W) can be calculated in the following way. The readout voltage from the array is $V(n)=n V_c$. The spread of switching currents for JJs in the array can be written as, $I_s(n)\simeq I_s^*+\sigma_n (n/N-1/2)$, where $I_s^*$ is the mean switching current and $\sigma_n$ is the width of the multi-junction switching current distribution. At the optimal bias, $I_b\simeq I_s^*$, the SCD response is $\Delta V \sim NV_c (I_{THz}/\sigma_n) $. 
Using Eq.(\ref{P_THz}), we obtain
\begin{equation}
S_n \simeq  \frac{2 \chi N V_c}{\sigma_n R_{THz} I_{THz}}.
\label{RespV}
\end{equation}

The procedure for quantitative estimation of sensitivity and noise-equivalent power (NEP) is described in Methods. For the case of Bi-2212 detectors from Figs. 5 and 6 we get $S>10^7$ (V/W) and $NEP\sim 10^{-17}~$(W$/\sqrt{Hz}$) for both MW and THz detection. There are very good numbers, however, far from optimal. Lets estimate achievable characteristics for a high-quality Bi-2212 mesa, like in Fig. 5 (b). Assuming $N=500$, $V_c=20$ mV, $I_s^*=50~\upmu$A; $\sigma_n=0.01 I_s^*$ (see Fig. 5 (c) at $T\sim 4$ K) 
\cite{Krasnov_2005,Krasnov_2024}); $R_{THz}=100~\Omega$ (for an efficient detector $R_{THz}$ should be comparable to the impedance of free space); and $I_{THz}=0.1 \sigma_n = 50~ $nA; 
we obtain $S \simeq 8 \cdot 10^{12}$ (V/W) and $NEP \simeq 0.1$ (zW$/\sqrt{Hz}$). This estimation is made under realistic assumptions and does not represent the ultimate limit of SCD \cite{Krasnov_2024}. 

Two factors contribute to this high sensitivity: the large cascade gain $\propto N$ and the narrow switching width $\sigma_n$. The latter depends on the uniformity of the array, inter-junction coupling, thermal fluctuations and electrical noise. When all junction in the stack are identical, they will switch at the same bias current. A spread of parameters cause broadening of $\sigma_n$. However, coupling between JJs leads 
to current-locking even of slightly dissimilar junctions  \cite{Grebenchuk_2022}. The true phase-locking of JJs can further reduce $\sigma_n$ \cite{Mros_1998}. Therefore, a strong inter-junction coupling is crucial for such a detector. Ultimately, $\sigma_n$ is limited by quantum fluctuations \cite{Martinis_1987,McDermott_2012,Andersen_2013,Inomata_2016,Ilichev_2017,Devyatov_1986,Krasnov_2005}, qualifying SCD as a quantum-limited photon detector. 

In conclusion, we have described the operational principle of a cascade switching current detector, utilizing arrays of coupled Josephson junctions. The coupling between junctions leads to the current locking phenomenon and results in avalanche-like switching of many junctions. This amplifies the readout voltage {\color{black} without changing the noise background and, therefore, enhances the signal-to-noise ratio and reduces NEP}. We have demonstrated two prototypes based either on low-$T_c$ linear Nb arrays, or high-$T_c$ stacked intrinsic Josephson junctions. Good impedance matching and absorption efficiency of the studied devices is confirmed by observation of geometry-specific antenna effects. We conclude that Bi-2212 whisker-based cascade SCD are promising candidates for broad-band and high-performance THz sensors. 

\section*{Methods}

{\color{black}
\subsection*{Experimental details}

Measurements are performed in a closed-cycle optical cryostat with a base temperature $\sim 3$ K at zero magnetic field. Arrays are current-biased using a programmable voltage source. The $I$-$V$ characteristics are measured in a quasi 4-probe configuration. Additional information about experimental setup, bias configuration and microwave measurements can be found in sec. SVII of the Supplementary. Details about fabrication and characterization of Nb arrays and Bi-2212 mesas are provided in sec. SVIII and SIX.

The operation of SCD requires statistical analysis. The switching statistics of Bi-2212 mesas for a single and multi-junction switching can be found in sec. SIV of the Supplementary. It is well described by conventional thermal-activation theory, described in sec. SII.  

The SCD response is measured via lock-in measurements. For this, a harmonic current, $I=I_b\sin(2\pi f_b t)$, with the amplitude $I_b$ and frequency $f_b=23$ Hz is supplied for a time interval of 1 s. The read-out voltage corresponds to the Fourier component of the detector $V(t)$ wave form at $f=f_b$. The lock-in response is calculated on-flight using FPGA-based FFT routine. Additional information about SCD operation and lock-in readout can be found in sec. SI and SV of the Supplementary.
} 

\subsection*{Quantitative estimation of detector characteristics}
For MW measurements the detector sensitivity is estimated as follows. First, the arriving MW power is measured using a cryostat as a bolometer. The sample stage in our cryostat has an effective thermal resistance $R_{th}\simeq 0.5$ K/mW \cite{Cattaneo_2021}. By measuring a small temperature rise, $\Delta T\sim$mK, caused by the MW beam we directly measure the total arriving power, $P_{tot}=\Delta T/R_{th}$. Because of the large $\lambda_0\simeq 4$ mm and significant diffraction at several narrow ($\sim 1$ cm) apertures of the cryostat, 
the power is fairly uniformly distributed over the sample space with the radius $r \sim 1.5$ cm and the mean power density $q_{MW}=P_{tot}/\pi r^2$. The incoming power on the device is then calculated as $P_{MW}=q_{MW}A$, assuming that $A=1~$mm$^2$ is the effective absorption area of the receiving antenna, formed by the electrodes. For the case of Fig. 4, $P_{tot} \simeq 10 ~\mu$W and $P_{MW} \simeq 14$ nW. 

The absorbed power, $P_a$, can be estimated from the suppression of the switching current, $\Delta I_s$, caused by the generated high-frequency current. In the MW case (low frequency limit) $\Delta I_s$ is just equal to $I_{MW}$ \cite{Krasnov_2024}. $P_a$ is calculated as \cite{Borodianskyi_2017}: $P_a \simeq (2\sqrt{2}/3\pi)(\Delta I_s/I_{c0})^{3/2}I_{c0} V_c$. For the case of Fig. 5 (d), $\Delta I_s \simeq 3~\mu$A at the attenuation 3.7 dB, $I_{c0}=38~\mu$A, $V_c=20~$mV, yielding $P_a\simeq 5$ nW, approximately 30\% of $P_{MW}$. The maximum voltage is $V\simeq 0.37$ V, and the sensitivity with respect to the absorbed power, $S_a=V/P_{a}\simeq 7.4~10^7$~(V/W) and the net sensitivity $S_{MW}=V/P_{MW} \simeq 2.6~10^7$~(V/W). 

For THz generation-detection test in Fig. 6 we can not confidently estimate the incoming power. However, the absorbed power can be estimated in the same way. For the green $I$-$V$ in Fig. 6 (e) at $V_{gen} =2.65$ V, $f_J=7.99$ THz, the suppression of switching current is $\Delta I_s=35~\mu$A. With $I_{c0}=385~\mu$A, $V_c=20~$mV and $V\simeq 1~$V it yields $P_a\simeq 60~$nW and $S_a\simeq 1.6~10^7$~(V/W). 

The noise-equivalent power is determined as $NEP=v_n/S$, where $v_n$ is the voltage noise. {\color{black} As shown in the Supplementary, in our setup it is in the range of few nV$/\sqrt{Hz}$, but could be reduced to below 1 nV$/\sqrt{Hz}$ using cryogenic preamplifiers.} Taking $v_n=1~$nV$/\sqrt{Hz}$ we obtain the following values of NEP with respect to absorbed power: $NEP_{MW}\simeq 1.35~10^{-17}$ W$/\sqrt{Hz}$ and $NEP_{THz}\simeq 6.25~10^{-17}$ W$/\sqrt{Hz}$.  

The achieved characteristics are very good, but far from optimal, as discussed above. The main reason for the larger $NEP_{THz}$ than $NEP_{MW}$ is that the mesa used for the THz test has a factor 10 larger $I_{c0}$. The NEP of SCD is approximately proportional to the Josephson energy, $E_{J0}$ \cite{Krasnov_2024}. Therefore, the weighted (by $E_{J0}$) THz sensitivity is actually twice better than for MW. 

\subsection*{Signal-to-noise ratio}

The signal-to-noise ratio in a cascade SCD is determined by the three quantities:

- The mean readout voltage $V_n$,

- The standard deviation (statistical uncertainty) $\delta V_n$,

- The noise threshold of readout electronics.

Direct noise measurements in our setup are presented in sec. SVI of the Supplementary. The noise floor is in the range 1-10 nV/Hz$^{1/2}$, limited by the intrinsic noise in the room temperature preamplifier. 

The effect of statistical uncertainty and the difference in sensitivities of single junction SCD and cascade SCD are discussed in sec. SIII. There we also write explicit equations for the sensitivity and provide corresponding numerical simulations. 

The bottom line of this analysis is the following: 
Cascading increases the mean readout $V_n$, generally improves the relative uncertainty, $\delta V_n/V_n$, and usually expands the dynamic range. But it does not affect the measurement noise (determined by the preamplifier). Therefore, cascading always increases the signal-to-noise ratio and reduces NEP, compared to a single-junction SCD, just as sketched in Fig. 1 (a).    

A good detector should have a high absorption efficiency of incident radiation. The maximum efficiency ($50\%$) is achieved at the impedance matching condition \cite{Balanis,Krasnov_2023}. However, a single JJ alone cannot provide proper matching because it is much smaller than $\lambda_0$ \cite{Krasnov_2023}. Therefore, the geometry of electrodes forming a receiving antenna, becomes crucial. In the case of linear Nb arrays, long electrodes function as traveling wave antennas, as evidenced by off-axis diagrams of both emission \cite{Galin_2018,Galin_2020} and absorption, Fig. \ref{fig:fig4} (c). 
Bi-2212 whisker mesas have a turnstile antenna geometry, enabling good impedance matching. This is evidenced by the high radiation efficiency ($12\%$ at $f \simeq 4$ THz) \cite{Cattaneo_2021} and the four-fold polarization diagram reported in Fig. \ref{fig:fig5} (f) \cite{Note1}. Thus, analysis of geometry-specific antenna effects provides a valuable tool for investigating impedance matching and the efficiency of both absorption and emission.
\\

\subsection*{Acknowledgements}

The work was supported by the Science for Peace and Security Programme, grant G5796. We are grateful to R. Gerdau and M. Galin for assistance with fabrication of Nb-arrays and to A. Kalenyuk 
for assistance with fabrication of Bi-2212 samples.


\end{document}